\font\mybb=msbm10 at 12pt
\def\bbxx#1{\hbox{\mybb#1}}
\def\Z {\bbxx{Z}}


\def \www{\Omega}

\def \ti {\tilde}

\def \2 {{1 \over 2}}
\def \3 {{1 \over 3}}
\def \4 {{1 \over 4}}
\def \5 {{1 \over 5}}
\def \6 {{1 \over 6}}
\def \7 {{1 \over 7}}
\def \8 {{1 \over 8}}
\def \9 {{1 \over 9}}
\def \0 { \infty}

\def\++ {{(+)}}
\def \- {{(-)}}
\def\+-{{(\pm)}}


\tolerance=10000
\input phyzzx

 \def\unit{\hbox to 3.3pt{\hskip1.3pt \vrule height 7pt width .4pt \hskip.7pt
\vrule height 7.85pt width .4pt \kern-2.4pt
\hrulefill \kern-3pt
\raise 4pt\hbox{\char'40}}}

\def\nup#1({Nucl.\ Phys.\  {\bf B#1}\ (}


\REF\Sag{A. Sagnotti,  in {\it Non-Perturbative Quantum Field Theory},
Proceedings
of
1987  Cargese Summer Institute, eds.
G. Mack et al  (Pergamon, 1988), p. 521.}
\REF\Hor{P. Ho\v rava,
Nucl. Phys. {\bf B327} (1989) 461;
Phys. Lett. {\bf B231} (1989) 251.}
\REF\Witten{E. Witten, Nucl. Phys. {\bf B443} (1995)  85, hep-th/9503124.}
\REF\Oopen{C.M.Hull, Phys. Lett. {\bf B357} (1995) 545,
hep-th/9506194.}
\REF\dab{A. Dabholkar, Phys. Lett. {\bf B357} (1995) 307, hep-th/9506160.}
\REF\PW{J. Polchinski and E. Witten, Nucl. Phys. {\bf B460} (1996) 525. }
\REF\GravDu{C.M. Hull, Nucl.Phys. {\bf B509 } (1998) 216, hep-th/9705162.}
 \REF\HT{C.M. Hull and P.K. Townsend, Nucl. Phys. {\bf B438} (1995) 109.}
\REF\Pol{J. Polchinski, ``Tasi lectures on D-branes", hep-th/9611050.}
\REF\GGP{G.W. Gibbons, M.B. Green and M.J. Perry, Phys. Lett.
 {\bf B370} (1996) 37.}
\REF\Doug{M. Douglas and M. Li, hep-th/9604041.}
\REF\HStr{C.M. Hull, Nucl. Phys. {\bf B468} (1996) 113, hep-th/9512181.}
\REF\Asp{P. Aspinwall, Nucl. Phys. Proc. Suppl. {\bf  46}
  (1996) 30, hep-th/9508154; J.H. Schwarz, hep-th/9508143.}
\REF\HW{P.~Ho\v rava and E.~Witten,
Nucl. Phys. {\bf B460} (1996) 506,
 {hep-th/9510209}
.}
\REF\appear{  E.~Bergshoeff, E.~Eyras, R.~Halbersma, J.P.~van der Schaar,
C.M.~Hull and Y.~Lozano,
perprint QMW-PH-98-39, UG-15/98, hep-th/9812.}
\REF\duf{M.J. Duff and J.X. Lu, Phys. Rev. Lett. 66 (1991) 1402; Nucl. Phys.
B357 (1991) 534; Class. Quant. Gr. 9 (1991) 1. }
\REF\di{J.A. Dixon, M.J. Duff and J.C. Plefka, Phys. Rev. Lett. 69 (1992) 3009.
}
\REF\isd{J.M. Izquierdo and P.K. Townsend, Nucl. Phys. B414 (1994) 93. }
\REF\bsdf{J. Blum and J.A. Harvey, Nucl. Phys. B416 (1994) 119.}
\REF\lsd{K. Lechner and M. Tonin,
Nucl. Phys. B475 (1996) 545.}
\REF\dufff{M.J. Duff, Class. Quant. Gr. 5 (1988) 189.}
\REF\s{A. Strominger, Nucl. Phys. B343 (1990) 167.}
\REF\chs{C.G. Callan, J.A. Harvey and A. Strominger, Nucl. Phys. B359 (1991)
611;
Nucl. Phys. B367 (1991) 60.}
\REF\witsmal{E. Witten, Nucl. Phys. {\bf B460} (1996) 541.}
\REF\anom{
J. Mourad,  Nucl.Phys. {\bf B512} (1998) 199-208, hep-th/9709012.}
\REF\dabsh{
A.~Dabholkar,
 hep-th/9804208.}



\Pubnum{ \vbox{ \hbox {QMW-98-43}   \hbox{hep-th/9812210}} }
\pubtype{}
\date{December, 1998}

\titlepage

\title {\bf  The Non-Perturbative $SO(32)$ Heterotic String}

\author{C.M. Hull}
\address{Physics Department,
Queen Mary and Westfield College,
\break
Mile End Road, London E1 4NS, U.K.}
\vskip 0.5cm

\abstract {
The $SO(32)$ heterotic string can be obtained from
the type IIB string by gauging a discrete symmetry that acts
as $(-1)^{F_L}$ on the perturbative string states and
reverses the parity of the D-string.
Consistency requires the presence of 32 NS 9-branes -- the S-duals of D9-branes
--
which give $SO(32)$ Chan-Paton factors to   open D-strings.
At finite string coupling, there are $SO(32)$ charges tethered to the
heterotic string world-sheet by open D-strings. At zero-coupling, the D-string
tension becomes
infinite and the $SO(32)$ charges are pulled onto the world-sheet, and give the
usual $SO(32)$
world-sheet currents of the heterotic string.   }

\endpage

The perturbative type I superstring can be obtained from the   perturbative
type IIB superstring by
orientifolding by the action  of the
world-sheet parity operator $\www$ [\Sag,\Hor]. However, it seems  that $\www$
extends to a
symmetry of the full non-perturbative IIB theory; indeed, it is straightforward
to extend its action
to BPS states and    to the dual IIB string theory that emerges in the strong
coupling
limit, where it acts through    a perturbative symmetry $\ti \www$ of the dual
theory. If $\www$
does extend  to such a non-perturbative symmetry, we can consider modding out
the IIB theory by this
symmetry for any value of the string coupling. In particular, in the strong
coupling limit, this
corresponds to modding out the dual IIB string theory by $\ti \www$,
and this should give the $SO(32)$ heterotic string, as this is the strong
coupling limit of the
type I string [\Witten-\PW].
This led to the conjecture of [\GravDu], that the
  $SO(32)$ heterotic string can be obtained from the     type IIB superstring
by modding out by the action  of the    symmetry $\ti \www$. The purpose of
this paper is
to invesitgate this construction further, and in particular show how the gauge
sector of the
heterotic string emerges.

In the orientifold construction of the type I superstring, it is essential to
add 32 D9-branes
which provide the Chan-Paton factors for the open strings.
Extrapolating to strong coupling, the D9-branes are replaced by their S-duals,
which are the NS-NS
9-branes proposed in [\GravDu]. Then   32 NS-NS 9-branes  are   required in the
construction of
the
$SO(32)$ heterotic string from the IIB string [\GravDu] and should play a key
role in
the construction of the  gauge sector.
We shall show that the 9-branes indeed give rise to the gauge sector. Moreover,
this
extends to a construction of the non-perturbative heterotic string and in
particular we will find
that at   finite string coupling, the $SO(32)$ charges are no longer confined
to the string
world-sheet but are tethered to it by strings, and as these strings  can break,
the charges can
escape.

The type IIB string theory has an $SL(2,\Z)$ U-duality symmetry [\HT], and the
massless bosonic fields   are $g_{MN}, B^1_{MN}, \Phi$ in the NS-NS sector and
$D_{MNPQ},  B^2_{MN}, \chi$ in the RR sector.
The 2-form fields $B^i_{MN}$ transform as doublets under $SL(2)$, while $SL(2)$
acts on
$\lambda
=
\chi +ie^{-\Phi}$   through fractional linear transformations.
There is also  a non-dynamical RR 10-form potential $A^2_{M...N}$ that couples
to D9-branes
[\PW]. This must fit into an $SL(2)$ doublet also [\GravDu], and its partner is
a
NS-NS 10-form potential $A^1_{M...N}$.
The theory has
D$p$-branes for $p=1,3,5,7,9$ [\Pol] and it is interesting to ask how these
transform under
$SL(2,\Z)$. The D3-brane is invariant,  but acting on the
D1, D5  and D9  branes  generates $(p,q)$-branes for all co-prime integers
$(p,q)$, as can be seen
from the superalgebra [\GravDu]. Of these
  1-branes, 5-branes and 9-branes,
the (0,1)-branes are D-branes, the (1,0)-branes couple to fields in  the NS-NS
sector, while the
$(p,q)$ branes are bound states of $p$ (1,0)  branes and $q$ (0,1) branes
(for   co-prime integers $(p,q)$).
The 1-branes and  5-branes couple to the 2-form fields $B^i_{MN}$ and the
9-branes can be thought of as coupling to  10-form potentials $A^i_{M...N}$.

The 7-brane is more subtle, as it couples to the scalar fields which transform
non-linearly under $SL(2,\Z)$. The
solutions of [\GGP], in which the axion ansatz
involves the modular invariant $j$-function, are $SL(2,\Z)$ invariant and
lead to singlet 7-brane charges.
Acting with $SL(2,\Z)$ on a 7-brane leaves its charge $Z_{i_1...i_7}$
invariant, but changes the
$SL(2,\Z)$ monodromy
and the couplings to
strings and 5-branes. These branes are characterised by two integers [\Doug]
and
are obtained from the $(0,1)$ D7-brane of
[\GGP] by an $SL(2,\Z)$ transformation;
we will refer here to
 the
7-branes  on which a
$(0,1)$ string can end  as $(1,0)$ 7-branes.

At weak coupling, $g \equiv <e^\Phi > \approx 0$, the perturbative states are
described by the  NS-NS or
(1,0) string   while all the other branes are non-perturbative [\HStr]. The
perturbative theory is formulated as the
usual type IIB superstring theory, with a topological expansion in terms of the
genus of the world-sheet of the
NS-NS string. At strong coupling, however, it is the RR or (0,1) string that
gives the states that are  perturbative
in an expansion in $\ti g = 1/g$. The   self-duality of the theory
implies that the strong-coupling
theory is again a type IIB string theory, but  now the formulation should be in
terms of the world-sheet of the
(0,1) string.
The perturbation theory in $\ti g$ is a sum over $(0,1)$ string world-sheets
with the power
of $\ti g$ corresponding to the genus of the world-sheet.
The dual perturbative string theory has left and right movers which each
decompose into an NS and
an R sector. For example, the fundamental string of the  weakly coupled
($g$-perturbative) theory
couples to the 2-form in the NS-NS sector but becomes the D-string of the dual
theory,
coupling to the the 2-form in the RR sector of the $\ti g$-perturbative
theory.

 At weak coupling, the (1,0) string can end on the D-branes carrying RR charge,
which are the 3-brane, and the (0,1) $p$-branes with $p=1,5,7,9$.
 In the  strongly coupled theory formulated in terms of the world-sheet of
the (0,1) string, the (0,1) string, the (0,1) 5-brane and the (0,1) 9-brane
carry charge that appears
in the NS-NS sector
of the dual (0,1) string   while the new D-branes with density proportional to
$1/\ti g$ on which
the (0,1) string can end are the
 3-brane, and the  (1,0) $p$-branes with $p=1,5,7,9$.
These dual D-branes all carry charge which occurs in the RR sector of the (0,1)
string. This
structure is obtained by
acting with the $SL(2,\Z)$ transformation
$$
S= \pmatrix {0 &1 \cr
-1 & 0}
\eqn\soi$$
which interchanges strong and weak coupling.
For example, this takes a (1,0) string ending on $(0,1)$ strings or 5-branes to
a (0,1) string ending on $(1,0)$ strings
or 5-branes.

The perturbative type IIB theory is formulated in terms of a fundamental
$(1,0)$ string and has a
symmetry $\www$ which reverses the parity of the $(1,0)$ world-sheet. The
dilaton is invariant under
$\www$, which  is a perturbative symmetry (i.e. it takes perturbative states to
perturbative states).
Of the massless bosonic fields, the ones that are invariant
under $\www$ are $g_{MN}, B^2_{MN},
\Phi, A^2_{M...N}$, while the others are odd,  and this tells us how it
acts on BPS states:
the  (0,1)  string, 5-brane and
9-brane are invariant while for the   (1,0)  string, the 2-form $B^1_{MN}$ to
which it couples is
odd, but the Wess-Zumino term in the world-sheet action is invariant   as
the world-sheet volume form is   odd under world-sheet parity
$\www$.
Similarly, the (1,0)
5-brane and
9-brane
together with the 7-brane and 3-brane
couple to $p+1$ form gauge potentials that are odd under $\www$, and so their
world-volume
volume-forms must also be odd under
$\www$, so that $\www$  must act as an orientation-reversing  world-volume
parity operator on these
branes.

The branes that couple only to the
invariant fields are     invariant, so that the only $(p,q)$
 D-branes  that are invariant are the D1,D5,D9  branes with charges   (0,1).
This tells us how to extend the action of $\www$ to the BPS brane sector, and
it will be assumed
that the perturbative symmetry extends to a symmetry of the full
non-perturbative type IIB string
(which will also be denoted $\www$).
This is the key assumption; if $\www$ did not extend to a symmetry of the full
non-perturbative
type IIB theory, then orientifolding by $\www$ would be problematic even for
small but finite
string coupling, whereas it is believed that the type I string extends to a
consistent non-perturbative theory, and this assumption is implicitly made in
many discussions of
string dualities. The existence of such a non-perturbative symmetry can be \lq
derived' from M-theory: the non-perturbative IIB theory arises from M-theory
compactified on a 2-torus in the limit in which both radii tend to zero [\Asp]
and the symmetry
that provides the   non-perturbative
version of $\Omega$
is given by the IIB theory limit of the  $\Z_2$ M-theory symmetry used by Ho\v
rava and Witten [\HW].

Then  $$ \ti \www = S\www S^{-1}
\eqn\tiw$$
is also a symmetry of the full IIB theory, where $S$ is the $SL(2,\Z)$
transformation \soi\
interchanging weak and strong coupling. However, $\ti \www $   leaves the
dilaton invariant, and is a  symmetry order by order in IIB string perturbation
theory.
The weakly coupled type IIB string   has a perturbative symmetry
$(-1)^{F_L}$, where $F_L$ is the
left-handed fermion number of the conventional world-sheet formulation,
and as $\ti \www  $ and $(-1)^{F_L}$ act in exactly the same way on
perturbative states,
$\ti \www  $ can be thought of as a non-perturbative extension of $(-1)^{F_L}$
[\GravDu].
 The massless bosonic fields that are invariant under  $\ti \www$ are $g_{MN},
B^1_{MN}, \Phi, A^1_{M...N}$, which are precisely the
  NS-NS fields, so that D-branes are odd and
NS or (1,0) strings, 5-branes and 9-branes are even under  $\ti \www$.
In the same way that $\www$ inverts the parity of the (1,0) string world-sheet,
together with the world-volume orientations of the
(1,0)
5-brane and
9-brane
and the 7-brane and 3-brane,
$\ti \www$   inverts  the parity of the (0,1) string world-sheet, and also
inverts the world-volume
orientations of the (0,1) 5-brane and 9-brane and the 7-brane and 3-brane.

The orientifold of the weakly coupled IIB theory
constructed using $\www$ gives the weakly coupled type I theory [\Sag,\Hor].
The orientifolding can be thought of as introducing an orientifold 9-plane, and
32 D9-branes must
be added to cancel the anomalies and divergencies  introduced by the
orientifold 9-plane.
  The invariant sector
is formulated in terms of unoriented closed type I
strings, with massless bosonic fields
 $g_{MN}, B^2_{MN}, \Phi$. In addition there is an open string sector, which
can (in some ways) be thought of as a
twisted sector, with $SO(32)$  Chan-Paton factors  arising from 32 RR 9-branes;
the Chan-Paton
factor labels which 9-brane the string ends on [\PW].
The RR string, 5-brane and
9-brane are invariant under
 $\www$, and so  survive the projection.
The closed fundamental IIB strings  are BPS,  and as $\www$ acts as world-sheet
parity on these,
they become the nonoriented fundamental   type I strings, which are
no longer BPS and can break.
As well   as non-oriented closed (1,0) strings, there should also be
non-oriented
(1,0) 5-branes, 3-branes and 7-branes  (and, in principle,   (1,0) 9-branes) in
the theory,
as $\www$ acts as an orientation-reversing parity operator on their
world-volumes.

An isomorphic construction emerges if we instead mod out the
strongly coupled IIB string using $\ti \www$. Indeed, the strongly coupled IIB
theory
is formulated as a world-sheet
theory of the (0,1) string, which is  a perturbation theory in $\ti g=1/g$,
and  $\ti \www$ reverses
the  parity of the (0,1) string world-sheet. The resulting type I theory is one
in which the
fundamental open and closed strings are (0,1) strings which can end on the
(1,0) strings, 5-branes and 9-branes. Consistency requires   32 (1,0) 9-branes
(which can be
thought of as cancelling out the effects of a (1,0) orientifold 9-plane), and
these give $SO(32)$
Chan-Paton factors to the open (0,1) strings. This is the standard type I
superstring, but embedded
differently in the type IIB theory. If this extrapolates to all values of the
coupling, then
continuing back to weak coupling, we find that modding out weakly-coupled
type IIB string by  $\ti
\www$ in the presence of 32 (1,0) 9-branes should give the weakly coupled
$SO(32)$  heterotic
string.  This construction of the $SO(32)$ heterotic string from the IIB
string modded out by $\tilde \Omega$ with 32 9-branes can be understood as a
particular limit of the Ho\v rava-Witten construction. In [\HW], M-theory on
$T^2$ is modded out by a $\Z_2$ symmetry to give M-theory on a cylinder
$S^1\times S^1/\Z_2$. In the limit in which the radius $R$ of the $S^1$ and the
length $L$ of the  $S^1/\Z_2$ both tend to zero,   with the limiting form of
$L/R$ small,  the M-theory on $T^2$ becomes the   IIB string with coupling
$L/R$, the $\Z_2 $ symmetry becomes $\tilde \Omega$ and the resulting theory is
the $SO(32)$ heterotic string.
Similarly, if $L/R$ is large, this reduces to the orientifold construction of
the type I string; see [\appear] for details.

This construction  implies that the gauge structure of the heterotic string
must arise
from the 32 (1,0) 9-branes. We will now show how this comes about, and verify
that the perturbative
heterotic string indeed emerges from this construction.

Consider first the D-string in the weakly-coupled type I string theory.
The zero-mode structure of the supergravity solution corresponding to the
D-string was studied in
[\Oopen,\dab] and shown to give precisely the world-sheet structure of the
heterotic string.
The excitation spectrum of the D-string can be calculated by
quantizing the open string   allowing Dirichlet (D) boundary conditions with
the fundamental
strings ending on the D-string, as well as Neumann (N) boundary conditions
[\PW].
There are three sectors: (i)  the NN sector in which both ends of the string
satisfy
Neumann boundary conditions (or, equivalently, end on the D9-branes) and carry
$SO(32)$ Chan-Paton
factors,
giving the usual
open strings of type I string theory; (ii)
the DD sector in which both ends  of the string lie on the D-string has a
massless sector
which is a world-sheet theory on the D-string, consisting of 8 scalars  and 8
right-handed
Majorana-Weyl world-sheet fermions; (iii)
the DN sector in which one end of the string lies on the D-string and the other
satisfies Neumann
boundary conditions and carries an $SO(32)$ Chan-Paton factor has a massless
sector which is again a
theory on the D-string world-sheet, this time consisting of 32
left-handed
Majorana-Weyl world-sheet fermions transforming as a {\bf 32} of $SO(32)$.
Thus the massless modes of the DD and DN sectors gives an effective D-string
world-sheet theory
 which is precisely that of the heterotic string [\PW].

Consider now the theory obtained by modding out the weakly coupled IIB string
by $\ti \www$, with 32
NS 9-branes. On the perturbative states, $\ti \www$ acts as $(-1)^{F_L}$ and
so,
of the massless fields in the IIB supergravity multiplet, an $N=1$ supergravity
supermultiplet
survives,
 with bosonic sector consisting of the NS-NS fields  $g_{MN},
B^1_{MN}, \Phi, A^1_{M...N}$.
The oriented fundamental string  and NS 5-brane of the IIB theory, coupling to
$B^1_{MN}$,
also survive, as do the 9-branes coupling to $A^1_{M...N}$.

As $\ti \www$ acts as the  parity operator on the IIB D-string world-sheet,
there are (0,1)
strings or \lq D-strings' of the orientifolded theory that are non-oriented and
can be closed or
open. The open D-strings can be thought of as ending on the NS 9-branes, and
these give them $SO(32)$
Chan-Paton factors.
They are not BPS states and  have interactions through which they can break and
join.
In the strong coupling limit, these D-strings become the fundamental open and
closed strings of the
dual type I theory, while in the weakly coupled theory, they have tensions of
order $1/g$ and
their effective dynamics is governed by a Born-Infeld action (without
Wess-Zumino term).
These D-strings can   end on the fundamental string and the solitonic 5-brane.
There are then DD D-strings which have both   ends lying on the fundamental
string or solitonic
5-brane, and DN D-strings, one end of which end of which is free and carries an
$SO(32)$ charge,
and the other end of which ends on the fundamental string or solitonic 5-brane.
There are thus
$SO(32)$ charges tethered to the fundamental string or solitonic 5-brane by
open D-strings.

In the zero coupling limit, the D-string tension becomes infinite, and these
D-strings collapse to
zero length. Then the $SO(32)$ charges tethered to the
fundamental string or solitonic 5-brane are pulled onto the world-volume,
giving rise to an $SO(32)$
current density on the world-volume.
As the D-strings collapse, all that survives is the zero-slope limit or
massless sector of the
D-string spectrum, and this must be the same as the massless sector of the
excitations of   type I strings ending on a D1 or D5-brane in the
weakly-coupled type I theory, as
the massless sector is in a short multiplet protected by supersymmetry that can
be extrapolated
from weak to strong coupling.
This means that for the D-strings ending on the fundamental string, all that
survives
are   8 scalars  and 8 right-handed
Majorana-Weyl world-sheet fermions from the DD sector and
32
left-handed
Majorana-Weyl world-sheet fermions transforming as a {\bf 32} of $SO(32)$ from
the DN sector.
This gives precisely the right world-sheet structure for the free fundamental
heterotic string.
What is novel is the picture of what happens to the heterotic string if the
coupling constant is
finite.
Then the  $SO(32)$  charges, which are confined to the world-sheet at zero
coupling,
can move off the world-sheet, but are tethered to it by strings of tension
$1/g$.
However, these can also break, so that a DN string splits into a DN string and
an NN string, which
is an open D-string that can then escape from the  fundamental string.

It has been realised for some time that the 5-brane in the $SO(32)$ theory
should carry $SO(32)$
currents [\duf-\chs], and in [\witsmal]   a   world-volume structure was
proposed, which was shown to lead to
 cancellation of all anomalies in [\anom]. For the D5-brane of the type I
theory, the massless
world-volume fields arise from type I strings with DD and DN boundary
conditions [\witsmal].
For $N$ coincident D5-branes,    the following 6-dimensional (1,0)
supermultiplets with $SO(4)$ R-symmetry arise.   The DD sector  gives a
 Yang-Mills multiplet with $Sp(N)$ gauge symmetry and a scalar multiplet
(transforming as an
$N(2N-1)-1$ antisymmetric tensor  representation of
$Sp(N)$, plus a singlet) with 4
scalars
$X$ and two spinors transforming as a vector and chiral spinor of $SO(4)$,
respectively;
the 4 $X^i$ are the collective coordinates for the brane.
The DN sector gives a pseudo-real hypermultiplet with 4 scalars which are a
vector of $SO(4)$ and a
4-component fermion which is an $SO(4)$ singlet.
These hypermultiplets transform as a $(32, d_a)$ of $SO(32)\times Sp(N)$ where
the representation of dimension $d_a=N(2N+1)$ is the adjoint of $Sp(N)$.

The same massless modes should be present on the heterotic 5-brane
world-volume, as the massless
sector can be extrapolated to strong type I (weak heterotic) coupling.
In the construction of the heterotic string by gauging $\ti \www$, the
world-sheet structure
emerges from D-strings ending on the heterotic 5-brane, and in the weak
coupling limit of the
heterotic string, the D-strings collapse to zero length and
only the massless sector survives.
The $SO(32)$ currents on the 5-brane world-sheet arise from $SO(32)$ charges
tethered to the
5-brane that are \lq pulled in' in the weak coupling limit, while the $Sp(2)$
currents arise from
non-oriented D-strings with both ends attached to the 5-brane.

The presence of the 9-branes plays a vital role in the construction and are
responsible for the
gauge structure. If no 9-branes are added, then the perturbative IIB string can
be orbifolded by
$(-1)^{F_L}$ to give the IIA string [\dabsh], and no gauge sector emerges.
 It is not clear whether this can
be extended to the full non-perturbative theory, and will be discussed further
in [\appear].

One can apply similar considerations to other string theories.
Consider the IIB theory. For weak coupling, the D-string dynamics is governed
by fundamental
strings ending on the D-string, which is useful as it allows detailed
calculation.
The D-string can be thought of as being surrounded by a cloud of fundamental
strings ending on
it, and   closed fundamental strings can break off from the tethered strings
and propagate into the
bulk, mediating the interaction with the bulk degrees of freedom. By duality,
at strong coupling the
(1,0) string dynamics is governed by (0,1) strings ending on the (1,0) string,
and this again can be
studied using (the S-dual) conformal field theory.
Back at weak IIB coupling, D-strings can end on fundamental strings, but this
is not so useful
in giving the dynamics of the fundamental strings, as we only have a
formulation of the D-string
dynamics in terms of the fundamental strings themselves (from which an
effective Born-Infeld action
can be derived).
However, the excitations of the D-string that are in short multiplets can be
understood, since they
can be extrapolated to strong coupling where they are the   short-multiplet
states of the fundamental
string. In particular, the massless sector of the the D-string is the same as
that of the
fundamental string, and consists of 8 world-sheet scalars $X^i$ and 16
world-sheet fermions, which
transform as  a vector and complex chiral spinor of the transverse $SO(8)$,
respectively.
At zero IIB string coupling, the tension of the D-strings becomes infinite and
only the massless
sector survives, so that the familiar world-sheet degrees of freedom of the
fundamental
IIB string arise from the D-strings attached to the fundamental string that
collapse to points,
giving local fields on the fundamental string world-sheet.
The picture suggests that at finite coupling, the world-sheet is replaced by
a cloud of D-strings and the non-BPS states might be understood in terms of
D-string excitations;
it would of course be very interesting to find a non-trivial check of this.

The description of the heterotic string proposed here arises from this picture
of the
type IIB string on gauging $\ti \www$. The D-strings become non-oriented and
non-BPS, but the
world-sheet structure of the free heterotic string again emerges from
D-strings attached to the world-sheet, in the limit in which they collapse to
points.

\refout
\bye